\documentclass[copyright]{eptcs} 
\providecommand{\event}{Post-Proceedings --- ThEdu'17} 
\usepackage{underscore}            

\usepackage[T1]{fontenc}
\usepackage{amssymb, amsmath}
\usepackage{graphicx}
\usepackage{varwidth}
\usepackage{tikz}
\usetikzlibrary{arrows,automata}
\usetikzlibrary{shapes.misc}
\newcommand\ownbreak[2][1.5cm]{\begin{varwidth}{#1}\centering#2\end{varwidth}}

\title{Learning how to Prove: From the Coq Proof Assistant to Textbook Style} 
\author{Sebastian B{\"o}hne \qquad \qquad Christoph Kreitz
\institute{Universit{\"a}t Potsdam\\
Institut f{\"u}r Informatik und Computational Science\\
August-Bebel-Stra{\ss}e 89\\
14482 Potsdam, Germany}
\email{\quad boehne@uni-potsdam.de \quad \qquad kreitz@cs.uni-potsdam.de}
}
\def\titlerunning{Learning how to Prove: From the Coq Proof Assistant to Textbook Style}
\def\authorrunning{Sebastian B{\"o}hne and Christoph Kreitz}

\begin{document}

\maketitle

\begin{abstract}

We have developed an alternative approach to teaching computer science students how to prove. First, students are taught how to prove theorems with the Coq proof assistant. In a second, more difficult, step students will transfer their acquired skills to the area of textbook proofs. In this article we present a realisation of the second step.

Proofs in Coq have a high degree of formality while textbook proofs have only a medium one. Therefore our key idea is to reduce the degree of formality from the level of Coq to textbook proofs in several small steps. For that purpose we introduce three proof styles between Coq and textbook proofs, called line by line comments, weakened line by line comments, and structure faithful proofs.

While this article is mostly conceptional we also report on experiences with putting our approach into practise.

\end{abstract}

\section{Introduction}

Most computer science students have difficulties with proving theorems. Since many of their solutions avoid formalisms or apply them in a wrong way (see~\cite{knobelsdorf2016}) it seems obvious that the formalisms are at the heart of the problem. In concreto, we suspect the blending of formal and more informal aspects of proofs to be the main obstacle to learning how to prove. The formal aspects make a precise argumentation necessary while the informal ones are hiding this precision.

In accordance with the Cognitive Apprenticeship approach~\cite{collins1991}, which we already used in another course (see~\cite{knobelsdorf2014}), we try to render the strategies for precise argumentation visible. Thus the key idea of our overall approach is to teach students to prove at a high formal level first and only then to transfer their achieved proficiencies (and not just the proofs) to the less formal but more usual textbook proof style.

Apart from transparency, focussing on the first step on formal proving has another -- perhaps even more important -- advantage: thanks to the symbiosis of formal proof style and computers we can use proof assistants that in turn provide the students with a clear and immediate feedback for every step they attempt. So strategies will be not only visible but even evaluable all the time. Therefore students can work on the first part under almost perfect learning conditions.

We were affirmed in this assessment by the results of a course, held in October 2016 at Universit{\"a}t Hamburg. In this course we tested the approach of the two steps with the focus being mainly on the first one. We decided to use the Coq proof assistant (see~\cite{inria} and~\cite{bertot2004}), since it provides the step by step feedback system we relied on.
Furthermore Coq has the advantage of being very flexible. This is due to the possibility of writing user defined tactics in Coq. In the next section we will illustrate how we made use of this possibility in the first step. Regarding the second step of transferring back to normal textbook proofs we had a rather naive approach: we gave a lot of assignments calling on the students to create a textbook proof out of a proof in Coq and vice versa. In practise, then, we spontaneously introduced a task of commenting the Coq code line by line as well. An elaborated version of this is a part of the approach presented in this article.

The results of the exam showed that students were doing very well on assignments related to the first step while they faced more and more problems the farther away the assignments were from the proof in Coq (see ~\cite[section 5.4]{knobelsdorf2017}).

We think there are at least three ways to deal with the difficulties the students had with the second step:
\begin{enumerate}
 \item Ignore them and focus on the ability to prove in Coq, because proof assistants are the future anyway.
 \item Change Coq proofs to be (more) like normal textbook proofs.
 \item Find some new way to bridge the gap between the ability to prove in Coq and in textbook style.
\end{enumerate}

We are sceptical about the first approach. There is a big difference between proving in prepared files and developing new projects independently in Coq, which includes finding the appropriate formalisations and notations, theorems, arrangement of them and so on. Furthermore, usually a lot of preparatory work has to be done to get to the point of interest.

The idea of the second approach would be that all of this preparatory work has to be anticipated and done before. So what is needed in this case are proof environments for the different domains of mathematics, which allow students to prove in Coq in a way similar to the textbook proof style.\footnote{
  This does not necessarily include the use of natural language, which is not the most important point of the intended resemblance.} 
In principle such environments can be developed in Coq by an excessive use of tactics and suitable libraries. So the second approach sounds very promising. Yet, such proof environments have to be realised and this is anything but trivial. It may be worth the effort but it may take several years, even when the respective domain is rather simple.

In this article we will concentrate on a realisation of the third approach, which we already tested in a course in September 2017 at the University of Potsdam. The key idea for dealing with the second step (the bridging) is to reduce the degree of formality stepwise from Coq to textbook style. We will describe this approach in more detail in section~\ref{sec:idea}. Before that we will present the Coq proof assistant in the way we use it and outline what the students are supposed to have learned after the first step: how can the students prove with the Coq proof assistant, what have they learned so far, what were the domains on which they learned it etc.~(section~\ref{sec:coq}).
In sections~\ref{sec:comments}~-~\ref{sec:structure} we will discuss the intermediate proof styles used for the stepwise reduction of the degree of formality. We call these line by line comments, weakened line by line comments, as well as structure faithful proofs. We will also discuss what the treatment of normal textbook proofs itself could look like in this approach (section~\ref{sec:assignments}).
Subsequently we will evaluate the concrete observations we made and the feedback we got from the students of the Potsdam course (section~\ref{sec:observations}). Although we do not have sufficient data yet to draw reliable conclusions we got some hints what to look for in the future. 
Section~\ref{sec:related} treats related work and section~\ref{sec:summary} provides summary and conclusion.

\newpage

\section{The First Step: Using the Coq Proof Assistant}\label{sec:coq}

The Coq proof assistant has been developed in the late 1980's by researchers at the French Institut national de recherche en informatique et en automatique (INRIA).
Like other proof assistants, Coq implements a higher-order type theory; thus theorems in Coq are understood as types and the proofs for theorems as elements of the respective type. This implies that the latter are lambda expressions and so the background of Coq is closely interwoven with functional programming.

However, one feature of Coq is that the user does not have to know or even be aware of all these issues. Instead the user can develop a proof step by step by applying so called tactics. Every step of the proof is evaluated for its correctness and its effects are shown to the user.

\begin{figure}[!b]
 \centering
 \includegraphics[keepaspectratio,width=\textwidth,height=\textheight]{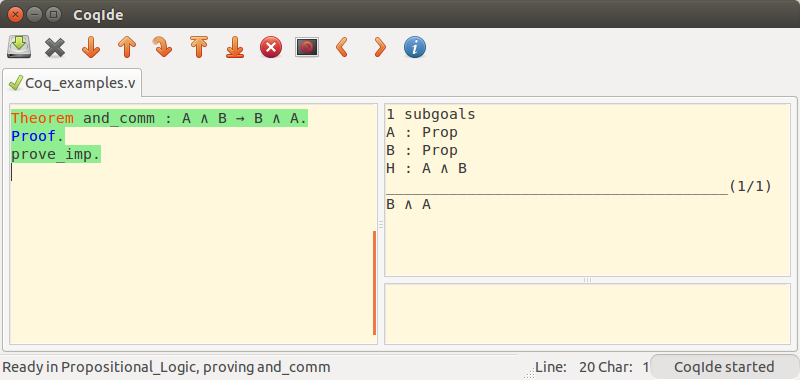}
 \caption{The proof of $A \wedge B \to B \wedge A$ after the first step.}
 \label{fig:andcomm1}
\end{figure}

\begin{figure}[!t]
 \centering
 \includegraphics[keepaspectratio,width=\textwidth,height=\textheight]{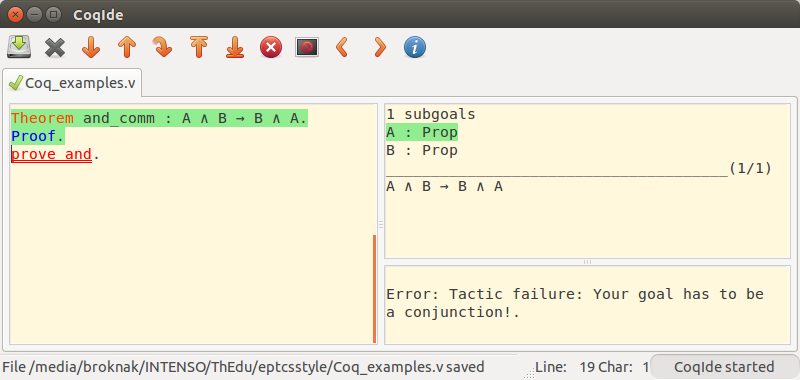}
 \caption{The proof of $A \wedge B \to B \wedge A$ after a wrong start.}
 \label{fig:andcomm2}
\end{figure}

\begin{figure}[!t]
 \centering
 \includegraphics[keepaspectratio,width=\textwidth,height=\textheight]{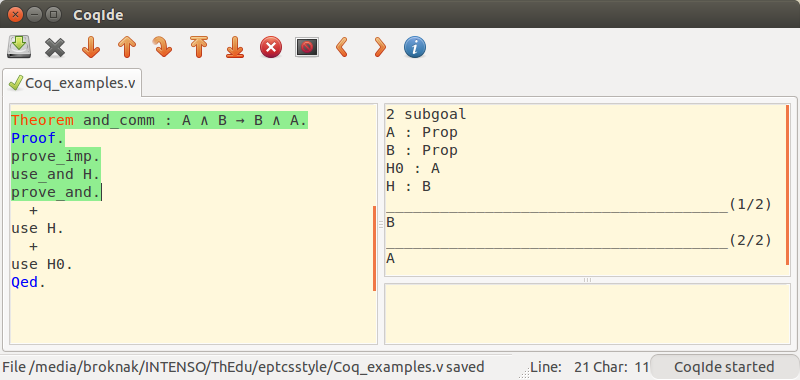}
 \caption{The complete proof of $A \wedge B \to B \wedge A$.}
 \label{fig:andcomm3}
\end{figure}

Let us illustrate this by a simple proof of $A \wedge B \to B \wedge A$ in the CoqIDE (see figure~\ref{fig:andcomm1}). We start by stating in the left of the window that we want to prove an implication. The Coq-System processes and shows this by greening our code, a clear positive feedback. Had we started by trying to prove a conjunction instead we would have received an error-message at the bottom right of the window (see figure~\ref{fig:andcomm2}).\footnote{
  This is idealised for the moment. In general suitable error messages have not been implemented yet.\newline~\newline}
In case we make a correct step the actual proof situation is shown at the top right of the window (figure~\ref{fig:andcomm1}). What we know (or assume) can be seen above the line while the goal(s) are listed below. If there is more than one goal then above the line only the context of the first subgoal is shown.

The remaining part of the proof is shown in figure~\ref{fig:andcomm3}. The screenshot shows the situation after step 3: $A \wedge B$ is already decomposed into $A$ and $B$ (via \texttt{use\_and}) and we have applied the tactic \texttt{prove\_and} to generate two subgoals, $B$ and $A$. The ``+'' in the next line focusses on the first of them, i.\,e.~after processing this symbol only the actual goal is shown. Both subproofs can be completed by using our assumptions.

\pagebreak

We have to mention that in Coq normally there are no tactics named \texttt{use\_xyz} or \texttt{prove\_xyz}. These tactics were defined by us and serve essentially as aliases for the logical tactics Coq is providing. We use the former for the following reasons:
\begin{itemize}
 \item Predefined Coq tactics have unstructured names and are therefore hard to remember; for tactic names such as \texttt{intro}, \texttt{case}, or \texttt{split} do not mention the relevant logical connectives and do not explain their relationship.
 \newpage	
 \item The names of standard Coq tactics do not make an explicit distinction in the treatment of assumptions and goals. By contrast, our tactics are called \texttt{use\_xyz} or \texttt{prove\_xyz}.
 \item Predefined Coq tactics have unwanted features. For instance \texttt{split}, the tactic corresponding to \texttt{prove\_and} also works on goals like $A \to B \to A \wedge B$.
\end{itemize}
So we designed and named the tactics in a way that requires students to reflect on how to prove theorems in Coq.	
\smallskip

After explaining the technical aspects of Coq let us now delineate what we wanted to teach the students with the help of this proof assistant. Our focus was mainly on what we call methodological aspects, i.\,e.~all aspects that matter for proving but are not related to specific contents or domains.
For instance, this includes logic, which in other mathematical courses is often buried by the contents and therefore not even recognised by the students anymore. We assigned great importance to it and the students had to start by proving theorems about propositional and predicate logic.
Other methodological skills needed were the proficiency to define data types and (non-recursive and recursive) functions, to unfold definitions at the right moment, and the knowledge how to prove theorems about inductively defined data types. 
In concreto we decided to treat simple contents such as seasons of the year (as an example for record types), lists, natural numbers, or binary trees for representing noninductive and inductive data types.

Figure~\ref{fig:successcoqproofs} shows the results of the course in Hamburg~2016. Students could handle the first step, i.\,e.~the formal theorem proving in Coq, already very successfully (see tasks~1,~2a,~3a,~3b,~4,~and~5a) while the results related to the second step were not equally convincing (tasks~2b,~5b,~and~6). Please note that there is a decreasing climax even within this second step: the further the task is away from Coq the worse the results. Hence in the following it will be the second step we focus our attention on.

However, before we proceed let us address the typical objection that the results of the exam related to the first step are rather deceiving. The idea behind this objection is that students allegedly could prove theorems in Coq by formal hacking without really mastering the level of Coq proofs. Yet, parts of the assignments during both courses -- especially in the domain of natural numbers -- were not solvable in this manner. Since students had no access to auto tactics the number of formal hacking attempts would be simply too big for some of the theorems there.

\begin{figure}[!b]
 \centering
 \begin{tabular}{|r|l|r|}\hline
  ~ & Task type & average of achieved \\
  ~ & ~ & points in percent\\\hline
  1 & Proving a propositional-logic-theorem in Coq & $100.00$\\\hline
  2a & Proving a first-order-logic-theorem in Coq & $90.34$\\\hline
  2b & Commenting this proof by natural language & $66.79$\\
  ~ & line by line & ~ \\\hline
  3a & Defining a nonrecursive data type in Coq & $100.00$\\\hline
  3b & Defining a function over this data type in Coq & $100.00$\\\hline
  4 & Proving a theorem about a given nonrecursive & $97.62$\\
  ~ & data type in Coq & ~ \\\hline
  5a & Proving an arithmetic theorem in Coq & $77.73$\\\hline
  5b & Developing a corresponding textbook proof & $53.23$\\\hline
  6 & Developing a textbook proof from scratch & $37.82$\\\hline
 \end{tabular}
 \caption{The results of the final exam for the Hamburg course 2016.}
 \label{fig:successcoqproofs}
\end{figure}

\newpage

\section{The Idea: Stepwise Reduction of the Degree of Formality}\label{sec:idea}

Before starting with the true content of this section let us introduce some terminology, which will be useful in this and the following sections. Instead of speaking of proofs of a high, medium, or low degree of formality we will sometimes just speak of high, medium, or low degree proofs. Note that high degree proofs in this terminology rather resemble programmes written in a low level language and vice versa. Furthermore, we will speak of the degree (of formality) of a proof style. By this we mean the respective degree of formality that all its proofs have. Since the platforms we will introduce later on are particular proof styles we will speak of the degree (of formality) of a platform, too.

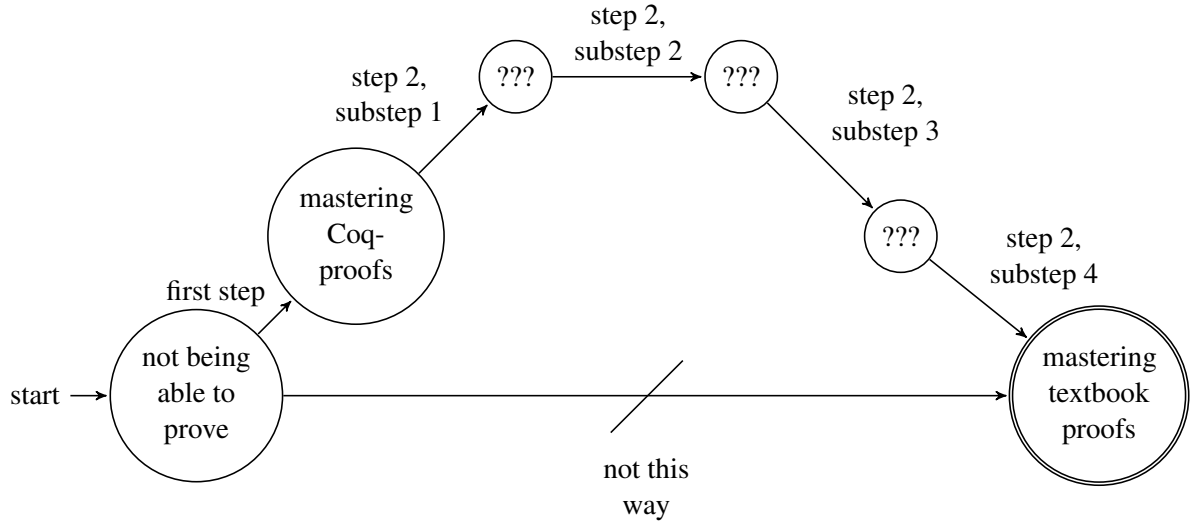
\begin{figure}
 \centering
  \begin{tikzpicture}[->,>=stealth', shorten >=1pt, auto, node distance=3.00cm,
                    semithick, every state/.style = {draw}]                  
  \node[initial,state] (A) {\ownbreak{not being able to prove}};
  \node[state] (B) [above right of = A] {\ownbreak{mastering Coq-proofs}};
  \node[state] (D) [above right of = B] {\ownbreak{???}};
  \node[state] (E) [right of = D] {\ownbreak{???}};
  \node[state] (F) [below right of = E] {\ownbreak{???}};
  \node (P1) [right of = A] {};
  \node[state] (P2) [-,strike out, right of = P1] {};
  \node (P3) [right of = P2] {};
  \node[accepting,state] (C) [right of = P3] {\ownbreak{mastering textbook proofs}};

  \draw[->] (A) -- node[below] {\ownbreak{~\\~\\not this way}} (C);
  \draw[->] (A) -- node {\ownbreak{first step}} (B);
  \draw[->] (B) -- node {\ownbreak{step 2, substep 1}} (D);
  \draw[->] (D) -- node {\ownbreak{step 2, substep 2}} (E);
  \draw[->] (E) -- node {\ownbreak{step 2, substep 3}} (F);
  \draw[->] (F) -- node {\ownbreak{step 2, substep 4}} (C);
 \end{tikzpicture}
 \caption{The first step is indeed a progress, but the major part is the former step two, which has to be divided in many different substeps.}
 \label{fig:intermediatesteps}
\end{figure}

The basic idea to improve our teaching regarding the second step, i.\,e.~the transferring to textbook proofs, is rather simple. We can compare this step with a stream
students have to cross. Some of them can jump across very well and are able to do so without further assistance. However, most of them will need help. Let us say there is a big flat stone in the middle. This would make things much easier. By contrast, if there would be a stone only at the beginning or at the end this would not be of much help. So we have to find helping stones dividing the whole step into approximately equidistant substeps (figure~\ref{fig:intermediatesteps}).
Such a scaffolding (and its later deinstallation) is in line with the Cognitive Apprenticeship approach (see~\cite{collins1991,knobelsdorf2014}).

What should the flat stones in the middle of the stream look like in the context of proving theorems? Our general approach is to teach students textbook proofs by teaching them proofs of a very high degree of formality first. So the idea to install platforms\footnote{
  In this article we use the word ``platform'' in a non-technical sense.}
being intermediate in its degrees of formality is not far fetched. Yet, the concrete choice is a little bit fiddly. The proofs one can develop with proof assistants like Isabelle/Isar or Mizar might be considered as appropriate regarding the degree of formality\footnote{
  We believe this to be a wrong assessment because Isabelle/Isar and Mizar proofs are rather a mix of high and low level proofs than medium degree proofs.}
\linebreak
(see~\cite[chapter\,2]{wenzel2017}) and with Back's structured derivations~\cite{back2010} there is even some pen and paper
\linebreak
approach. Yet, these approaches are product oriented and therefore do not fit our process oriented approach (see section~\ref{sec:related} for further discussion). Therefore we had to invent new suitable proof styles.

One clue to do so is language itself. Experience from both courses shows that students face problems in transferring from the formal language to natural language (and vice versa). Furthermore in some cases we even observed reluctance towards natural language; perhaps (those) students thought of natural language as less precise and transparent. Therefore intermediate platforms should stress that natural language can be used in a very precise manner, too. Furthermore students have to learn that natural language has the power to structure arguments. For instance, language allows to emphasize the main branch of a proof by putting the others in ``since \ldots'' clauses~etc. and so language can linearise a proof. Furthermore language can structure arguments by providing explicit orientation.

Another decisive point is condensation in all its forms of appearance. In Coq every single step, i.\,e.~modification of the actual proof situation, is explicitly mentioned. In addition every step has an explicit justification (the applied tactic with its arguments). If analogous situations appear the respective argumentation has to be executed repeatedly. All this is not the case in textbook proofs. Hence we have to find intermediate platforms regarding condensation as well.

The last point we want to mention is readability. Textbook proofs provide us with a guiding thread. They illuminate what is going on and explain on a rather intuitive level why this works. Although even in textbook proofs we find lots of technical details, compared to the Coq proofs treated so far they are rather given from a bird's than a worm's eye perspective. So readability should be one aspect of our partitioning of the second step.

In the next three sections we will present intermediate platforms pursuant to the above considerations about language, condensation, and readability. Since we started with a special form of high degree proofs, namely the ones in Coq, we can (and will) use the presentation of proofs specific to Coq to define the intermediate platforms. Furthermore we will discuss for each intermediate platform to what extent feedback is possible.

\section{A Small Step for Mathematicians but a Big One for Learners: Line by Line Comments}\label{sec:comments}

\enlargethispage{\baselineskip}

\begin{figure}
 \centering
 \includegraphics[keepaspectratio, height=0.8\textheight]{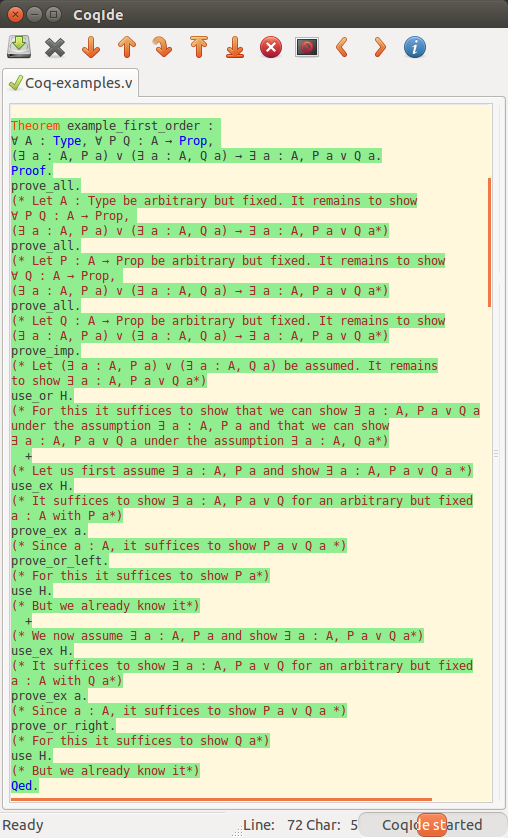}
 \caption{A proof exemplifying the line by line comments}
 \label{fig:examplefirstorder}
\end{figure}

The first intermediate platform we introduce is \emph{line by line commentation}. By this we mean adding comments to every line of an already existing Coq proof. These comments have to use natural language to describe precisely the modifications made in the proof situation at this step. This is illustrated in figure~\ref{fig:examplefirstorder}. To support this tasks students are provided with formulations for all the logical rules and the remaining tactics that can appear.

Please note regarding language the sole focus is on precision. Hence (for now) we do not request the students to vary their formulations. Even more we do not expect the students to use language in a structure giving way or an ideas mediating manner. Furthermore at this stage we do not require any form of condensation, whatsoever.

Another relevant point is that students need some time to write down comments. Since the writing process itself does not exhaust their mental resources, students will also reflect on what they are doing during their process. Furthermore they have to make explicit the bookkeeping that is usually done automatically by Coq. Hence they do not only have to handle every new situation for itself anymore, but they have to view their proof holistically. So every step becomes part of a whole and the different proof step situations become linked together. Hence we do foster the thinking about the Coq proof at hand and therefore line by line commentation is also a tentative step towards a bird's eye perspective.

\pagebreak

Obviously, we do not have any kind of automatic feedback for the comments. Yet, this does not mean that there is no feedback at all. Due to the proximity of the comments between the Coq code and due to the given recommendations for formulations students can create their own feedback. By contrast, when proving a theorem in a textbook manner there are no such clear reference points for self-monitoring.

\enlargethispage{\baselineskip}

To sum up, we have created a Coq dependent proof style, which is between the Coq proof itself and textbook variants of it. Since it is the first of three intermediate platforms this style is still closer to the former than to the latter. Yet, this level already involves language and some transition to a bird's eye perspective. Furthermore the feedback is already weakened. That weakening can indeed be a step forward will be the subject of the next section.

\section{Weakened Line by Line Comments}\label{sec:weakened}

By a \emph{weakened line by line commentation} we mean a line by line commentation with a few important differences:
\begin{enumerate}
 \item There are some situations where two or more steps of the same kind must be condensed into a single step. For example the first four steps in figure~\ref{fig:examplefirstorder} can be condensed to ``Let $A$ be a type, $P$ and $Q$ be propositional functions over $A$, and let furthermore $(\exists~a:A,~P~a)~\vee~(\exists~a:A,~Q~a)$ be assumed. We have to show $\exists~a:A,~P~a~\vee~Q~a$''.
 \item There is some textual smoothing required. The previous point already contains the formulations ``be a type'' and ``propositional function over A''. Furthermore from this level on many textual repetitions should be avoided.\footnote{
   In the course in Potsdam we did not require to avoid repetitions but it seems to be a suitable point.}
 \item Trivial endings like ``but we already know it'', ``the same are always equal'', or ``ex falso quodlibet''~etc. are not commented anymore. Instead the comments for these are subsumed in the respective comments before (if there are any).
 \item The comments have to be oriented at an intuitive understanding instead of the logical rules. In figure~\ref{fig:examplefirstorder}, for instance, we simplify the comment after ``use\_or'' by ``We make a case analysis over $(\exists~a:A,~P~a)~\vee~(\exists~a:A,~Q~a)$''.
 \item In case of analogous branches in the proof only the first of them is permitted to be commented. The following ones have to be declared as analogous. Furthermore it has to be explained precisely what changes would have to be made if we had commented everything. In contrast to the proof in figure~\ref{fig:examplefirstorder}, for instance, a weakened line by line commentation would only write one comment for the second branch: ``This case is analogous to the first one. $P~a$ has to be replaced by $Q~a$ everywhere except for the disjunction $P~a~\vee~Q~a$ and we have to prove the right side of the disjunction instead of the left one''.
\end{enumerate}

The use of language becomes more difficult at this stage. First of all the formulations must have a richer variety now. This in turn eliminates the possibility of a complete listing of all needed formulations, which was given to the students in the case of line by line commentation before. So students cannot simply replicate the suggested formulations from such a given list but have to modify the former formulations. In case of analogous branches, students even have to find formulations completely on their own. Finally, some of the formulations must condense several lines. So language starts to determine the structure of the proof. Yet, this is just the beginning and most parts are still given by the Coq proof. Furthermore language is still not responsible for the communication of ideas.

Despite the lack of ideas contained in them the weakened line by line comments become more readable. This is due to the more intuitive formulations and the omission of trivial endings. Furthermore the labelling of branches as analogous saves resources.

The latter point deserves further attention. When writing (or reading) line by line comments students should become frustrated by all the annoying repetitions. They should wish to condense these and this happens at this stage for the first time. This pertains to the analogous branches as well as to lines of the same kind. Yet, both kinds of condensation preserve information. They can be reverted and are therefore innocuous. 

Regarding feedback comparison with canned formulations will not always be possible anymore. Instead students have to orient themselves on other weakened line by line comments or their former solutions for line by line comments. Although this does not sound like much it is still much more than in the case of textbook proofs. In the former only narrow deviations from a given standard have to be found.

The weakening of line by line comments has brought us closer to textbook proofs. While the main focus in the transition to line by line comments is on language the focus in the transition to weakened line by line comments is on condensation. Although we already made some tentative improvements regarding readability there is still a big gap between weakened line by line comments and textbook proofs. With the next intermediate platform we will reduce it.

\section{Structure Faithful Proofs}\label{sec:structure}

The last of the three intermediate platforms is given by what we call \emph{structure faithful proofs}. These are proofs in natural language that have the same structure as the corresponding Coq proofs. Besides the usual backward reasoning structure faithful proofs allow forward reasoning using equations and implications. Except for argumentations based on reflexivity, symmetry, and transitivity of equality or other equivalence relations every single step has to be mentioned to some extent. For every step a justification has to be stated but this should be a short hint only. For instance, theorems have to be mentioned but not the arguments with which they are instantiated. Furthermore there is an important consequence of this definition: since structure faithful proofs do only depend on the structure of Coq proofs but not on the lines of code, they can be developed without having to write down a Coq proof.

Let us consider an example of a structure faithful proof (figure~\ref{fig:structurefaithfulproof}). What comes to our attention immediately is the explicit structure given by ``+'', ``*'', and ``-''. Apart from that the proof looks similar to a textbook proof. Most of the justifications in this proof can be found on top of the equality signs. That justifications usually are only hints in this kind of proof can be seen, for example, in $\textit{Suc~} n \ominus \textit{Suc~} l \stackrel{\{\textit{suc\_n\_sub\_suc\_m}\}}{=} n \ominus l$. Here, we express that the theorem \texttt{suc\_n\_sub\_suc\_m} is used but we do not state explicitly that $n$ from the theorem is instantiated with $n$ and $m$ with $l$. This does not only pertain to equational reasoning: ``[\ldots] we have $n \ominus l  = 0$ and therefore by equ\_fct, $\textit{pred~} (n \ominus l) = \textit{pred~} 0$'' is another example for this less precise kind of argumentation.

\begin{figure}
 \centering
 \parbox{0.87\textwidth}{
 Theorem: $\forall~n~m : \mathbb{N},~n \ominus m \not= 0 \vee n = m \to \textit{Suc~} n \ominus m = \textit{Suc~} (n \ominus m)$.\\
 Proof: Let $n$, $m$ be arbitrary but fixed natural numbers and let us assume $n \ominus m \not= 0 \vee n = m$. We have to prove $\textit{Suc~} n \ominus m = \textit{Suc~} (n \ominus m)$. We do this by case analysis of the disjunction.
 \begin{enumerate}
  \item[$+$] We assume $n \ominus m \not= 0$ and show $\textit{Suc~} n \ominus m = \textit{Suc~} (n \ominus m)$. By the structure theorem of $\mathbb{N}$ we know $m = 0 \vee (\exists~l : \mathbb{N},~m = \textit{Suc~} l)$. So we can prove our goal by another case analysis.
  \begin{enumerate}
   \item[$*$] We assume $m = 0$. Then we do have $\textit{Suc~} n \ominus m \stackrel{\{m~=~0\}}{=} \textit{Suc~} n \ominus 0 \stackrel{\{\textit{n\_sub\_$0$}\}}{=} \textit{Suc~} n$ on the left hand side and $\textit{Suc~} (n \ominus m) \stackrel{\{m~=~0\}}{=} \textit{Suc~} (n \ominus 0) \stackrel{\{\textit{n\_sub\_$0$}\}}{=} \textit{Suc~} n$ on the right hand side. So we have the same on both sides.
   \item[$*$] We assume $m = \textit{Suc~} l$ for some $l : \mathbb{N}$. This delivers $\textit{Suc~} n \ominus m \stackrel{\{m~=~\textit{Suc~} l\}}{=} \textit{Suc~} n \ominus \textit{Suc~} l \stackrel{\{\textit{suc\_n\_sub\_suc\_m}\}}{=} n \ominus l$ on the left hand side. On the right hand side we have $\textit{Suc~}(n\ominus m) = \textit{Suc~}(n\ominus \textit{Suc~}l) \stackrel{\{\textit{n\_sub\_suc\_m}\}}{=} \textit{Suc~} (\textit{pred~} (n \ominus l))$.
   \begin{enumerate}
    \item[$-$] If we could prove $n \ominus l  \not= 0$, we would have furthermore $\textit{Suc~} (\textit{pred~} (n \ominus l)) \stackrel{\{\textit{suc\_pred\_n}\}}{=} n \ominus l$, such that left and right hand side would be equal.
    \item[$-$] So it remains to show that $n \ominus l  \not= 0$. For this we assume $n \ominus l  = 0$ and derive a contradiction. On the one hand we do have $0 \not= n \ominus m \stackrel{\{m = \textit{Suc~}l\}}{=} n \ominus \textit{Suc~}l \stackrel{\{\textit{n\_sub\_suc\_m}\}}{=} \textit{pred~} (n \ominus l)$. On the other hand we have $n \ominus l  = 0$ and therefore by equ\_fct, $\textit{pred~} (n \ominus l) = \textit{pred~} 0 \stackrel{\{\textit{pred\_$0$}\}}{=} 0$, the required contradiction. 
   \end{enumerate}
  \end{enumerate}
  \item[$+$] We assume $n = m$ and show $\textit{Suc~} n \ominus m = \textit{Suc~} (n \ominus m)$. On the left hand side we have $\textit{Suc~} n \ominus m \stackrel{\{n = m\}}{=} \textit{Suc~} n \ominus n \stackrel{\{I\_add\_n\}}{=} (1 \oplus n) \ominus n \stackrel{\{\text{add\_comm}\}}{=} (n \oplus 1) \ominus n \stackrel{\{\textit{n\_add\_m\_sub\_n}\}}{=} 1$. The right hand side delivers $\textit{Suc~} (n \ominus m) \stackrel{\{n = m\}}{=} \textit{Suc~} (n \ominus n) \stackrel{\{\textit{n\_sub\_n}\}}{=} \textit{Suc~} 0 \stackrel{\{\textit{by Def}\}}{=} 1$. \hspace*{0mm}\hfill q.e.d.
 \end{enumerate}
 }
 \caption{Example of a structure faithful proof.}
 \label{fig:structurefaithfulproof}
\end{figure}

The reader should now have an impression of what a structure faithful proof looks like. So let us now start to analyse this kind of proof.
Regarding language we are close to the goal now. In this respect the only remaining difference to textbook proofs is that language is not used to structure the proof per se. In particular we are still not able to give more weight to some of the branches. However, within the subproofs language already is used freely and therefore structures the argument. So the use of language is still a big hurdle for students.

\newpage

The equational reasoning is a new way to condense a lot of steps. Instead of giving each equality a justification before or after we put the equations together and the justifications above the equations. So far nothing spoils restorability. Yet, the justifications themselves are only hints now and so some information is really lost.

What sounds like a drawback is actually a feature since omitting details increases readability. The addition of forward elements helps to create a guiding thread, which again increases readability. Finally, the same is true for the flexible use of language. For instance, we can accentuate focussing on the essential parts of a subproof.

There is only a single rather external kind of feedback left. This is the structure the students can orient themselves on. They can see explicitly where they are in their proof and therefore it is easier to keep an overview of what is given at the moment and what is to do next. In addition to that in some parts of the proof students can orient themselves on the formulations used in (weakened) line by line comments. Yet, in most situations there is no unique or at least a standard formulation given to orient oneself on.

We have now seen an intermediate platform that has its main focus on the aspect of readability. Except for the explicit structure, with its forward elements and its reduced demand for precision, it reminds one already of textbook proofs. However, the explicit structure is still a big support since the requirement to prioritise different branches to make the proof more linear is not easy to fulfil.

In our comparison with the crossing of a river the next step leads to firm ground, namely the textbook proofs. They are what we intended to teach the students. Let it be mentioned, however, that in principle we do not have to stop at textbook proofs since they are not the least formal kind of argumentation. We can conceive them as only one further platform in the stream, the other end of which could be the proof ideas. In this case we would propose one further intermediate platform between textbook proofs and proof ideas. The corresponding proof style gives the technical details of the essential parts of the proof (like in the textbook proofs) but omits the details of other parts.
In our teaching the key word to differentiate between textbook proofs and this variant is ``show'' instead of ``prove''. Yet, in practise there does not seem to be a clear distinction with respect to the use of both words.

\section{How to Teach It?}\label{sec:assignments}

Until now we have presented the different platforms (line by line comments, weakened line by line comments, and structure faithful proofs). In this section we will focus on the steps between these platforms; i.e.~we try to answer how students can master some stage when the previous ones have already been learned.

In general we propose to spend most of the time with supervised training sessions where the students work on their own. Though autonomous working on assignments is definitely not a new idea it is of utmost importance in this case: the focus is not on contents but on methods and these can be learned best by applying them. The role of the lecturer should be limited to a brief introduction of the platform per se, including the presentation of solutions to some examples before the students start to work on their own. The lecturer should moderate the discussion of student's solutions only where appropriate.

So what should assignments for students look like? Of course, they should mirror the transitions. Therefore a lot of assignments require the students to take a solution of stage $n$ and to transfer it to stage $n + 1$. For a better grasp of the relation between two such stages students are asked to transfer a few solutions of stage $n +1$ back to stage $n$. 

\newpage

Yet, in the end students are not supposed to start with a Coq proof and transfer it step by step to textbook proofs but they should be able to develop the latter directly. 
So first, distance must be broadened, i.\,e.~we recommend also assignments that require the transferring from stage $n$ to stage $m$ where neither $n$ nor $m$ is the immediate successor of the other. Naturally, the farther the treatment of an intermediate platform $m$ is the farther the distance between $n$ and $m$ can be. In particular the ending point might be a textbook proof.
Second, students must get rid of the Coq crutch entirely. So we need assignments that do not presuppose any solution of a previous stage but instead ask the students to develop a proof of the respective stage from scratch.

However, in its character these assignments differ from the other ones since there is no given structure the students can orient themselves on. Therefore these kinds of exercises should be divided further gradually: the first assignments of this kind should be to prove theorems which are clear with respect to structure. For instance, theorems using the pumping lemma are good candidates here, whereas the latter assignments should be allowed to conceal this structure, like most theorems in fact do. Proving that a set is enumerable, for instance, belongs to that kind of assignment. In such cases students must be trained to find the structure (including the relevant types) to apply their transition skills from there on.

\section{Observations, Feedback, and Exam Results}\label{sec:observations}

In our first attempt to teach students textbook proofs via Coq proofs we were in a luxurious situation regarding evaluation. We had two experts for collecting data, who could spend a lot of their time on that matter during the whole course (see~\cite{knobelsdorf2017}).
Since the main focus of this course was on the first step, learning Coq proofs, the results are unfortunately limited with respect to the second step, the transition from Coq proofs to textbook proofs.

This time we had no resources for such a professional investigation. Nevertheless we collected data. They stem from
\begin{enumerate}
 \item feedback forms the students were asked to fill in twice during the course.\footnote{
   We had 9 of 12 students filling in the first feedback form and 8 students filling in the second.}
 Mainly these consisted for each intermediate platform of questions regarding fun, success and estimated impact for learning textbook proofs.
 \item observations, which due to the intensive supervision and interaction with the students should be much more precise than in a traditional lecture.
 \item the final exam.\footnote{
   It was written by 9 students.}
\end{enumerate}
Although we do not have sufficient data yet to draw reliable conclusions they already show some tendencies and indications.

Let us briefly discuss the background of the students of our course. According to the feedback the students in general were slightly above average in their respective course of studies. However, their performances were widely spread. Three of them estimated themselves to be among the top $10\%$ of their class, while one believed to be in the bottom $10\%$. Our observations confirmed this wide spread. Apart from two exceptions most of the students had been exposed to proofs rather frequently. Although there were students convinced to belong to the best $10\%$ no one rated his skills in understanding and developing proofs at the highest level. In average students estimated their skills here to be between medium and weak with understanding being closer to medium and developing being closer to weak. Students rated the usability of Coq as good. So difficulties should not have arisen from that aspect.

The first point we want to discuss is the fun aspect. In our first attempt to teach textbook proofs via Coq proofs in Hamburg we had recognised that students had fun developing proofs in Coq. However, we also had recognised some unwillingness to switch back to natural language proofs. Hence before the second course in Potsdam we were uncertain how students would digest the much longer period of dealing with proofs in natural language. Furthermore we feared the line by line commentation and its weak version to be so monotonous that students could lose all their motivation at these stages. However, the feedbacks indicate that the effect is not that dramatic, though it seems to exist. While the fun aspect for proving in Coq was ranked between medium and high, the fun aspect of the intermediate stages was ranked a little bit below medium (except for structure faithful proofs, which were slightly above medium). Surprisingly the rating for developing textbook proofs was only at the same level as the intermediate stages.
The students' feedback is in accord with our observations. Students worked focussed in all their assignments for intermediate steps. There was no grumbling. In the case of weakened line by line comments two students were questioning the repetitions but there was no questioning of the use of natural language or one of the intermediate platforms per se.

With respect to performance the exam results should have the highest validity but we also have the feedback and observations at our disposal. All but one of the assignments that ended in Coq code have very good results (see figure~\ref{fig:successtextbookproofs}, assignments 1a, 2a, 3, 4a, 4b and the exception~6). Students were equally successful regarding the line by line commentation (assignment~1b). The result with respect to weakened line by line commentation is not as convincing (assignment~2b) but the errors made there were often related to a confusion of the different platforms. So it does not seem to be the platform per se with which the students struggle. This is also suggested by the feedback where students rated their success in line by line commentation and its weakened version even higher than in the case of Coq proofs. There was no assignment requiring the development of a structure faithful proof (only one starting from such a proof) but the feedback and our observations indicate the performances there to be between an average and decent level.

\begin{figure}[!b]
 \centering
 \begin{tabular}{|r|l|r|}\hline
  ~ & Task type & average of achieved \\
  ~ & ~ & points in percent\\\hline
  1a & Proving a first-order-logic theorem in Coq & $98,41$\\\hline
  1b & Developing corresponding line by line comments & $97,04$\\\hline
  2a & Changing the code of a Coq proof according to some criteria & $86,51$\\\hline  
  2b & Developing corresponding weakened line by line comments & $66,36$\\\hline
  3 & Transferring a structure faithful proof to a Coq proof & $97,69$\\\hline
  4a & Defining a recursive function in Coq & $86,67$\\\hline
  4b & Proving an arithmetic theorem in Coq & $83,80$\\\hline
  4c & Developing a corresponding textbook proof & $55,98$\\\hline
  5 & Developing a textbook proof about binary trees from scratch & $52,96$\\\hline
  6 & Developing a Coq proof using arithmetic tactics & $44,97$\\\hline  
 \end{tabular}
 \caption{The results of the final exam for the Potsdam course 2017.}
 \label{fig:successtextbookproofs}
\end{figure}

However, the performances regarding the intermediate platforms are not too interesting per se. What is decisive is their potential to help to learn textbook proofs. For evaluation of this we can adduce the feedback and the exam results. Note that we do not mention observations here since unfortunately at the end of the course there were time conflicts with exams of other courses and so most students were not present in the training sessions.

The feedback given by the students seems to be inconsistent. On the one hand the usefulness of the single intermediate platforms with respect to textbook proofs was assessed between medium and high. To be more specific, the structure faithful proofs were rated best (close to high), followed by line by line comments (between medium and high), and the weakened version of them (close to medium). On the other hand the students gave an overall rating of the usefulness of the course. It turned out that students assessed it only between low and medium. A few students criticised that there was still too big a gap between structure faithful proofs and textbook proofs, because the theorems of the former were already formalised or had a clear formalisation while some of the latter had not.

Let us compare this now with the exam results. Assignment~4, which required a textbook proof, but only after finding a Coq proof first, was the same as assignment~5 in the Hamburg course (see figure~\ref{fig:successcoqproofs} on page~\pageref{fig:successcoqproofs}).\footnote{
  Part a of the Potsdam exam, the definition, was already provided by us in the Hamburg exam.}
The results regarding this textbook proof were minimally above average in both cases. This, of course, is not too good news for our approach. However, with respect to assignment~5 (assignment~6 of the Hamburg course), in which a proof had to be developed from scratch, the picture changes: while in Hamburg the students reached only $38$\% in average, in Potsdam this increased to $53$\%. Of course, this can still be improved but at least it indicates a clear progress. 

On top of that we think that the results regarding textbook proofs could be misleading. Assignment~6 should have been at least close to the other assignments done in Coq. Yet, the results were much worse. We see no other explanation for this than the time conflicts with other exams at the end of this course. But if this is true then the same pertains to the assignments requiring the development of textbook proofs. In other words, when there was already a progress under poor conditions the advance under good conditions might turn out too be much bigger.

Whatever the progress of our approach may truly be, the solutions of the relevant assignments show clearly that it is not divided equally. Despite our effort a few students were not even able to start a solution. We can exclude time problems within the exam since the respective students finished their exams early. Furthermore some students may have been doing well without our course being responsible for this. However, more than the half of the exams bear the hallmark of this course and -- except for one case -- the respective solutions are good. So it seems that we have found a way to improve proving skills for many but not all the students. We are planning future investigations to have more reliable and more differentiated data: which types of students react positively to our approach and what can be done to help the others?

\section{Discussion of Related Work}\label{sec:related}

In visual appearance our different proof styles, including the Coq proofs, resemble the structured derivations presented by Back in~\cite{back2010}. For instance, we find indentations and justifications similar to ours.  Particularly strong is the resemblance in the case of structure faithful proofs since structured derivations make use of equational reasoning, too, and the same kind of justifications (like transitivity) are left implicit. With us Back's approach shares the idea of transparency on the teaching and the learning side.  However, there is no process of different proof styles in Back's case. Furthermore the main difference to our approach is that the whole idea of structured derivations
is product-oriented, while we focus on the proof process.

\enlargethispage{\baselineskip}

There is a variety of approaches using proof assistants for teaching. Several discuss the poor performances of students in proving~\cite{hoover1996, nipkow2012, retel2005, sakowicz2007}. Nipkow calls the student's creations in~\cite{nipkow2012} LSD trip proofs. Hoover and Rudnicki~\cite{hoover1996} stress that poor proofs can be found even in textbooks. We only want to add that the inability to create solutions at all might be the biggest problem.

Many authors agree that feedback is of utmost importance~\cite{carter2013, hoover1996, narboux2005, nipkow2012, retel2005, sakowicz2007, trybulec1993}\footnote{
  In the first case the term ``interactivity'' is used instead.}
The most traditional view is that feedback assures us what (parts of) proofs are correct~\cite{narboux2005, retel2005, trybulec1993}. Nipkow~\cite{nipkow2012} stresses gamification aspects, namely that feedback is addictive and serves as a challenge that causes students to work a lot harder. Like us, he criticises that conventional homework corrections have too low a frequency. Carter and Monk~\cite{carter2013} require a frequent, immediate, and clear feedback in the learning process. We would like to point out that the last attribute cannot be accomplished in full if proof assistants employ an automatic justification mechanism, as this may permit steps too big or reject steps that should be fine but cannot be handled by the system. This point is conceded in articles on Mizar~\cite{hoover1996,trybulec1993}. However, what is completely new in our article is the idea to reduce the amount of feedback during the course, which should cause students to mature and become independent of feedback eventually.

Many articles attach great importance to using a tool that does not require students to learn too much of the system before teaching the true contents of the course. While Nipkow~\cite{nipkow2012} stresses that only a quarter of the time of a course was devoted to teaching Isabelle/HOL, all Mizar articles~(\cite{hoover1996, retel2005, trybulec1993}) concede that learning how to write proofs with Mizar is challenging and requires much effort in courses. Others modify a standard system to avoid problems with the system itself. The web server ProofWeb~\cite{hendriks2010} was developed to avoid installation issues and problems with different versions of Coq. The Papuq system~\cite{sakowicz2007} is a modification of Coq that doesn't require students to learn the original Coq tactics. Instead in every proof situation it shows a window with natural language suggestions how to proceed. The Lurch system~\cite{carter2013} goes even further, since it resembles a word processor and is easy to use even for non computer affine students. By contrast, we did not modify the Coq proof assistant itself but only added tactics.  We want to concentrate on the conceptional aspects of the course and for that the original CoqIDE is good enough, as confirmed by the students surveys.

The work of~\cite{hendriks2010}~and~\cite{hoover1996} is restricted to logic. Sakowicz and Chrz{\k{a}}szcz~\cite{sakowicz2007} treat a simplified version of
set theory in their course. Algebra and number theory is taught with Lurch on top of logic and set theory in~\cite{carter2013}. Trybulec and Rudnicki~\cite{trybulec1993} concentrate on relations while Nipkow~\cite{nipkow2012} uses the Isabelle/HOL to teach semantics. We devoted the non logical part to data structures but could imagine treating other parts, especially relations, as well in future courses.

In recent times using proof script based proof assistants for teaching has fallen into disrepute. Instead it is endorsed to use declarative proof assistants like Isabelle/Isar (see~\cite{wenzel2017}) or Mizar, in which the proofs are of a synthetic nature and only the new knowledge is stated step by step while the justification is done (mostly) by auto tactics. Nipkow~\cite[section\,3.3]{nipkow2012} criticises that proof script based proof assistants are not readable by humans, lack structure, do not make ideas visible, and -- most importantly -- have a relation to real proofs that can be compared to the relation between assemblers and normal programming languages. While the first point may be true, the second
and third depend mostly on the person developing the proof, and the comparison in the fourth is flawed.  The implicit argument here is that one does not have to learn assemblers first to learn programming and in the same way one does not have to learn proving in a proof script based proof assistant first. But procedural thinking is firmly fixed in human beings while static truths are not. Hence more students will have an intuitive grasp of programming than of proving. Thus students have to learn the basic principles of proving first such that they become second nature to them. Having said that, however, we do consider the use of declarative proof assistants as promising whenever the students have already learned how to prove.

Especially Hoover and Rudnicki in~\cite{hoover1996} stress a method called structuring of proofs. By this they mean the introduction of indented blocks having assumptions at the beginning, some steps in between, and a result at the end. Blocks are evaluated as a whole in the ongoing proof. An implication, for instance, is shown by such a block where the assumption is the premise and the result is the conclusion of the block. This is related to our structuring in Coq and the intermediate steps but comprises many more cases. All the examples given in~\cite[section\,A.5]{hoover1996}, for instance, do not require any structure in our case. This is because  our approach is (initially) analytic while theirs is synthetic.

What is missing in all cited articles -- except for Carter and Monks'~\cite{carter2013}, where students develop proofs with Latex in the end -- is a treatment of the relation between the kind of proofs done in a proof assistant and the more informal textbook proofs. One reason is that the focus is often on machine readable proofs and not on transferring these to textbook proofs. In our opinion the difficulty and the value of transferring the skill of proving in a proof assistant to the skill of textbook proving is extremely underestimated.

\section{Summary and Conclusion}\label{sec:summary}

Most computer science students have difficulties with proving theorems. We have developed an approach in which students learn to prove in Coq first and start to develop textbook proofs only after that. In this article we focussed on teaching the transition to textbook proofs. The main idea was to find intermediate proof styles that reduce the degree of formality step by step. To do so we emphasized three aspects of a degree of formality: language, condensation, and readability. Furthermore feedback, which is not part of a degree of formality itself but a related consequence, was involved in our considerations.

We started with a concrete instance of high degree proofs, namely proofs in Coq, and defined new proof styles that depended on this. In concreto, we presented line by line comments, their weakened version, and structure faithful proofs. Each of these was characterised by a leap in one of the three aspects: language in the first, condensation in the second, and readability in the last case. In addition, there were some changes in the remaining aspects. We also described types of assignments that can lead to success in textbook proving.

Finally, we analysed the data we got from feedback, observations, and exam results. The data show that students do well with the intermediate platforms and the corresponding assignments. The results regarding the final, decisive transition to textbook proofs indicate a clear progress but not a panacea for all the problems related to the development of textbook proofs. One further improvement for the future could be to stress the formalisation of theorems even more in order to give the students a better understanding of the relation between formal and informal formulations of theorems.

From a bird's eye perspective our data indicate that -- at least for computer science students -- formalisms per se are not the problem in learning textbook proofs (but rather the mix of formal and less formal elements) and that it is a good approach to start proving with a suitable proof assistant even if one only wants to teach textbook proofs. We expect this to hold for students of mathematics as well. For a substantiation of this claim we aim at extending our empirical base to gain reliable data in the future.
\smallskip

This article is part of a larger investigation on different degrees of formality. Most of the other parts are of a more philosophical nature and pertain to questions about the characterisation and the achievement of different degrees of formality. However, there is one other part that deals with Coq. It builds on the idea to bring Coq proofs close(r) to textbook proofs via proof environments, as described in the introduction, that go beyond currently available systems. Such proof environments should be an alternative to face the problems with the second step. We would simply spirit it away. Furthermore such proof environments would be very interesting for experienced human provers, too. Proofs developed this way could convey ideas and would have the same persuasiveness for humans as textbook proofs while providing the same assurance as any other verified proof.

\end{document}